\documentclass[11pt]{article}
\usepackage{amssymb,amsmath,cite,geometry,graphicx,url,blkarray,tikz,hyperref}
\usepackage[normalem]{ulem}
\usepackage{color}
\tikzstyle{every node}=[circle, draw, fill=black!50,
inner sep=0pt, minimum width=4pt]

\catcode`\@=11

\@addtoreset{equation}{section}

\long\def\@makefntext#1{\parindent 0cm\noindent
\hbox to 1em{\hss$^{\@thefnmark}$}#1}

\newcommand{\captionfonts}{\small}
\makeatletter  % Allow the use of @ in command names
\long\def\@makecaption#1#2{%
  \vskip\abovecaptionskip
  \sbox\@tempboxa{{\captionfonts #1: #2}}%
  \ifdim \wd\@tempboxa >\hsize
    {\captionfonts #1: #2\par}
  \else
    \hbox to\hsize{\hfil\box\@tempboxa\hfil}%
  \fi
  \vskip\belowcaptionskip}
\makeatother   % Cancel the effect of \makeatletter 

\begin{document}
\begin{titlepage}
\vspace{.5in}
\begin{flushright}
July 2022\\  %date
\end{flushright}
\vspace{.5in}
\begin{center}
{\Large\bf
 Path integral suppression of badly behaved causal sets }\\  %title
\vspace{.4in}
{P.~C{\sc arlip}}\footnote{\it email: peter.carlip@gmail.com}\\
        {\small\it unaffiliated}\\[.4ex]
{S.~C{\sc arlip}\footnote{\it email: carlip@physics.ucdavis.edu\\\hspace*{.85em}ORCID iD 0000-0002-9666-384X}\\
        {\small\it Department of Physics}\\
       {\small\it University of California}\\
       {\small\it Davis, CA 95616, USA}}\\[.4ex]
     {S.~S{\sc urya}} \footnote{\it email: ssurya@rri.res.in}\\
        {\small\it  Raman Research Institute, }\\ {\small \it CV Raman Ave, Sadashivanagar} \\ {\small \it  Bangalore, 560080, India} \\
     
\end{center}

\vspace{.5in}
\begin{center}
{\large\bf Abstract}
\end{center}
\begin{center}
\begin{minipage}{4.75in}
{\small
Causal set theory is a discrete model of spacetime that retains a notion of causal structure.  We understand 
how to construct causal sets that approximate a given spacetime, but most causal sets are not at all 
manifold-like, and must be {{{dynamically excluded} }} if something like our universe is to emerge from the theory.
Here we show that the most common of these ``bad'' causal sets, the Kleitman-Rothschild 
orders, are strongly suppressed in the gravitational path integral, and we provide evidence that a large 
class of other ``bad'' causal sets are similarly suppressed.  It thus becomes plausible that continuum 
behavior could emerge naturally from causal set quantum theory.}
\end{minipage}
\end{center}
\end{titlepage}
\addtocounter{footnote}{-1}

\section{Introduction}

To build a discrete model of spacetime, one must first choose which aspects of spacetime
structure to keep.  In causal set theory, the choice is the causal structure, the relativistic
picture of time in which each event has a past, present, and future.  A causal set is essentially a
discrete, causally ordered collection of events, with an added prohibition of closed timelike curves 
\cite{causets}.

The appeal of this approach comes in part from theorems of Hawking, King, McCarthy, and Malament
\cite{Hawking,Malament},
who show that if an ordinary continuum spacetime obeys suitable causality conditions, its geometry is
completely determined by its causal structure and volume element.  A causal set embodies the
discrete version of this information: the causal structure is built in, and the volume of any region is
proportional to the number of points it contains.  

This is not quite enough, though; we must still ask how well the discrete structure approximates 
the continuum.  There are actually two questions:
\begin{enumerate}
\item Given a spacetime, can we construct a causal set that approximates it at some scale, and use
that causal set to reconstruct the properties of the spacetime?
\item If we start with an arbitrary causal set, can we find a spacetime it approximates?
\end{enumerate}

A good deal is known about the first question.  Given a spacetime $M$, there is a well understood procedure,
``Poisson sprinkling,''  for producing a causal set that approximates $M$.  From such a causal set, we can 
reconstruct the dimension of $M$, its coarse-grained topology and geometry, and such structures as 
d'Alembertians and Greens functions.  While important open issues remain, the answer to question 1 is 
largely ``yes.''

For question 2, { much less is known, and on the face of it the situation
  seems dire}.  Most causal sets fail to approximate any manifold
at all.  In fact, if one chooses a large random causal set, one is overwhelmingly likely to find a 
Kleitman-Rothschild (KR) order, a ``three-layer'' set with large spatial extension but only three moments 
of time \cite{kr}.  If one discards the KR orders, one is overwhelmingly likely to be left with a two-layer set, followed 
by an infinite sequence of higher layered but still non-manifoldlike sets {\cite{dhar,pst}}.  If causal sets are fundamental, 
something profound must happen to eliminate these ``bad'' sets and leave us with an approximate continuum.

A first step towards addressing this problem was taken in \cite{Loomis}, where it was shown that for a
large range of coupling constants, the gravitational path integral very strongly suppresses the two-layer
causal sets.  In \cite{Surya}, this result was extended to a larger class of multi-layer sets, including the KR orders.  
To show this, though, the authors of \cite{Surya} had to use {a} ``wrong'' action, {which is } a truncated version of the full 
Einstein-Hilbert action.  

Here, we complete the proof for KR orders, showing that they are superexponentially suppressed in the 
gravitational path integral with the full Einstein-Hilbert action. {In a subsequent work we will explore this question
  for the subset of all {higher layered} causal sets that are sub-dominant compared to the KR
  orders \cite{dhar,pst}.}   While this does not settle the question---there are certainly other types
 of ``bad'' non-manifoldlike causal sets---these results makes it much more plausible that our observed continuum 
 spacetime could emerge from a discrete causal set path integral.

 \section{Causal sets}

 A causal set is a locally finite, irreflexive partially ordered set. The
partial ordering is interpreted as causal ordering: $x\prec y$ means
``$x$ is to the causal past of $y$,'' where the irreflexive condition
forbids  $x \prec x$.    Local finiteness means that for any $x$
and $y$, the number of points $z$ for which $x\prec z\prec y$  is
finite; this is a discreteness condition.   Irreflexivity implies that
there are no points $x,y$  for which both $x\prec y$ and $y\prec x$
(if there were, this would imply $x\prec x$); this is a causality
condition, excluding  closed causal  curves.

We will need several causal set structures that have continuum analogs.
\begin{itemize}
\item[--] A chain is an ordered set of elements $x_1\prec x_2\prec\dots\prec x_n$.  This is an analog of a causal path.
\item[--] An antichain is a set of unrelated elements $\{x_1,x_2,\dots,x_n\}$ (that is, $x_i\nprec x_j$ for any $i,j$).
This is an analog of a spacelike hypersurface.
\item[--] Given two points $x,y$ with $x\prec y$, the order interval $I(x,y)$ is the set of points $z$ such that 
$x\prec z\prec y$.  This is an analog of a causal diamond or Alexandrov open set, the region within the
future light cone of $x$ and the past light cone of $y$.   
\item[--] A pair $x\prec y$ for which $I(x,y)$ is empty---that is, 
a pair of ``nearest neighbors''---is called a link.  Note that linked points are ``near'' in the Lorentzian sense:
their \emph{proper} distance is small, but they may have large separations {in
  coordinate space or time}.
 \end{itemize}
 
 Given a finite volume region $M$ of a causal spacetime of dimension $d$, there is a standard way to construct a corresponding
 causal set.  In a Poisson sprinkling, points from $M$ are chosen randomly and independently at a density
$\rho$, with causal relations induced from the causal structure of the manifold.  One then ``forgets'' the manifold, 
keeping only the selected points and their causal relations, which naturally form a causal set.  The density $\rho$ 
determines a characteristic discreteness scale $\ell {=\rho^{-\frac{1}{d}}}$, below which the causal set contains little information about 
$M$.  But on scales large compared to $\ell$,  the dimension and 
much of {the}  topology and geometry {of $(M,g)$ can be reconstructed from the causal set \cite{lr}}.  

Most causal sets, however, are nothing like these manifold-like sets.  If one picks a random causal set with $n$ 
points, for $n$ large enough one is overwhelmingly likely to obtain a Kleitman-Rothschild (KR) order \cite{kr}.  This
is a  causal set with three layers (see Fig \ref{kr:fig}), in which
\begin{itemize}
\item[--] layer 1 and layer 3 each have $\frac{n}{4} + {\mathcal{O}(n^{1/2} \log n)}$ elements 
\item[--] layer 2 has $\frac{n}{2} + {\mathcal{O}}(\log n)$ elements
\item[--] each element in layer 1 and layer 3 links with $\frac{n}{4} + {\mathcal{O}}(n^{7/8})$ elements in layer 2 
\item[--]  each element in layer 2 links with $\frac{n}{8} + {\mathcal{O}}(n^{7/8})$ elements in layer 1 and in layer 3
\end{itemize}
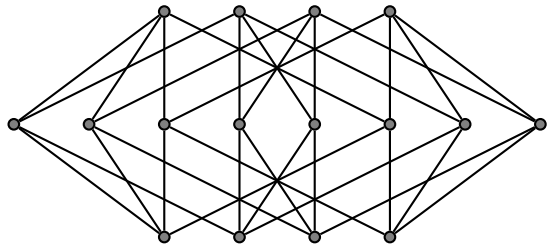
\begin{figure}
\begin{center}
\begin{tikzpicture} [style=thick] 
  \node (a1) at (0,0){};
  \node (a2) at (1,0) {};
  \node (a3) at (2,0) {};
  \node (a4) at (3,0) {};
  
  \node (a5) at (-2,1.5) {};
  \node (a6) at (-1,1.5) {};
  \node (a7) at (0,1.5) {};
  \node (a8) at (1,1.5) {};
  \node (a9) at (2,1.5) {};
  \node (a10) at (3,1.5) {};
\node (a11) at (4,1.5) {}; 
\node (a12) at (5,1.5) {};

 \node (a13) at (0,3) {};
 \node (a14) at (1,3) {}; 
 \node (a15) at (2,3) {}; 
 \node (a16) at (3,3)   {};

  \draw (a1) -- (a5);
  \draw (a1) -- (a6);
  \draw (a1) -- (a7);  
  \draw (a1) -- (a10);
  \draw (a2) -- (a5);
  \draw (a2) -- (a8);
  \draw (a2) -- (a9);
  \draw (a2) -- (a11);  
  \draw (a3) -- (a6);  
  \draw (a3) -- (a8);
  \draw (a3) -- (a9);
  \draw (a3) -- (a12);
  \draw (a4) -- (a7);  
  \draw (a4) -- (a10);
  \draw (a4) -- (a11);
  \draw (a4) -- (a12);
 \draw (a13) -- (a5);
  \draw (a13) -- (a6);
  \draw (a13) -- (a7);  
  \draw (a13) -- (a10);
  \draw (a14) -- (a5);
  \draw (a14) -- (a8);
  \draw (a14) -- (a9);
  \draw (a14) -- (a11);  
  \draw (a15) -- (a6);  
  \draw (a15) -- (a8);
  \draw (a15) -- (a9);
  \draw (a15) -- (a12);
  \draw (a16) -- (a7);  
  \draw (a16) -- (a10);
  \draw (a16) -- (a11);
  \draw (a16) -- (a12);
 
\end{tikzpicture}
\caption{An example of a KR order.} \label{kr:fig} \end{center}
\end{figure} 

{The number of $n$-element causal sets grows as $\sim 2^{\frac{n^2}{4}+\frac{3}{2}n+o(n)}$. By ``overwhelmingly
likely'' we  mean that the ratio \cite{kr} 
$$\frac{\hbox{number of $n$-element KR orders}}{\hbox{number of all $n$-element causal sets}} \sim 1
+O\biggl(\frac{1}{n}\biggr) , $$
i.e., the number of $n$-element KR orders also grows as $\sim 2^{\frac{n^2}{4}+\frac{3}{2}n+o(n)}$.}

The full picture is even worse.  If the KR orders are removed by hand, the remaining sets are overwhelmingly
likely to be two-layer sets.  If those are removed, one is left with four-layer sets, {followed by}  an infinite
hierarchy of higher layered sets \cite{dhar,pst}. The ``nice'' manifoldlike causal sets obtained from Poisson sprinklings 
are apparently vanishingly rare.

\section{Path integrals and the Benincasa-Dowker action}

The fact that ``physical'' causal sets are so rare may be a cause for concern, but it is not such a strange
situation.  Even in a system as simple as a free point particle, the space of smooth paths is of
measure zero.  In that case, we understand what happens: the ``bad'' paths are suppressed by destructive
interference in the path integral, leaving only a tiny subset of physical trajectories.  It is natural to ask whether
the same happens for causal sets.

To answer this, we would ideally construct a causal set action from first principles and then evaluate the
path integral (technically, in view of discreteness, the path sum).  We do not know how to do this.  We can, 
however, use our knowledge of manifoldlike causal sets obtained from Poisson sprinklings to construct
the causal set analog of the Einstein-Hilbert action.

For a causal set $C$ with $n$ elements, let $N_J$ be the number of $J$-element order intervals,\footnote{Note
that conventions vary in the literature;  our $N_J$ is sometimes called $N_{J+1}$} that is, the number of 
pairs $x\prec y$ for which there are exactly $J$ elements $z$ such that $x\prec z\prec y$.  $N_0$ is the number 
of links, and has a maximum value {$N_0^{\rm max} =\frac{n^2}{4}$ }.  As Benincasa and Dowker have
shown, for causal sets obtained from a Poisson sprinkling of a four-dimensional spacetime $M$, the 
Einstein-Hilbert action is then well approximated by the quantity
\begin{align}
 \frac{I_{\scalebox{.65}{\em BD}}(C)}{\hbar} ={2\sqrt{\frac{2}{3}}}\left(\frac{\ell}{\ell_p}\right)^2(n - N_0 + 9N_1 -16N_2 + 8N_3) ,
\label{b1}
\end{align}
where $\ell_p$ is the Planck length and $\ell$ is a discretization scale \cite{bd}.  More precisely, if one
takes the expectation value of (\ref{b1}) over Poisson sprinklings, the difference between that expectation
value and the Einstein-Hilbert action for $M$ goes to zero as the sprinkling density $\rho$ goes to infinity.
Intuitively, (\ref{b1}) counts causal diamonds of varying volumes, and those volumes depend on curvature;
the alternating signs are needed to cancel contributions from points that are ``near'' in  proper
distance but lie far down the light cone.

We can now write the path sum 
\begin{align}
\mathcal{Z}({\Omega}) = \sum_{C\in {\Omega}}\exp\left\{\frac{i}{\hbar}I_{\scalebox{.65}{\em BD}}(C)\right\}
\label{b2} 
\end{align}
over some collection
$\Omega$ of  { $n$-element} causal sets, and ask whether it is suppressed.  This is not easy, since the
number of distinct {$n$-element} causal sets increases as {$2^{\frac{n^2}{4}+\frac{3}{2}n+o(n)}$}.  But as shown in \cite{Loomis},
if we take $\Omega$ to be the set of two-layer causal sets, the sum (\ref{b2}) can be evaluated.  The
simplification comes from the fact that in a two-layer set, the only order intervals are links, so the action (\ref{b1})
depends only {on} $n$ and $N_0$.  A combinatoric argument then tightly bounds the number of sets with a given 
value of $p=N_0/N_0^{\rm max}$.  For large $n$, the sum (\ref{b2}) can then be approximated by an integral 
over $p$ and evaluated by steepest descent.  The result is that for
\begin{align}
\tan\left[{\sqrt{\frac{2}{3}}}\left(\frac{\ell}{\ell_p}\right)^2\right] > \left(\frac{27}{4}e^{-1/2}-1\right)^{1/2}
\label{b3}
\end{align}
$\mathcal{Z}(\mathcal{C}_{2-layer})$ goes as $2^{-cn^2}$ for some positive constant $c$.  If the discreteness
occurs near the Planck scale, this  is a truly enormous suppression: a spacetime region of one cubic centimeter 
times one nanosecond already contains $n\sim10^{133}$ Planck volumes.

The two-layer sets are the second most common ``bad'' causal sets, so this is a hopeful sign.  As a next
step, these results were partially generalized in \cite{Surya} to the KR orders, the most common ``bad'' sets.   To do so, though, the authors of \cite{Surya} had to restrict themselves to the ``link action''
\begin{align}
 \frac{I_{\scalebox{.65}{\em link}}(C)}{\hbar} = {2\sqrt{\frac{2}{3}}}\left(\frac{\ell}{\ell_p}\right)^2(n - N_0) 
\label{b4}
\end{align}
rather than the full Benincasa-Dowker action (\ref{b1}).  The action then again depends only on the ratio
$p=N_0/N_0^{\rm max}$, and a similar combinatoric argument bounds the number of sets with a given value of $p$.
In the end, one obtains exactly the same condition (\ref{b3}) on the couplings, and the same superexponential
suppression.{This is already very interesting since it suggests that the calculation of \cite{Loomis} for bi-layer
  orders contains a crucial nugget relevant to all $k$-layer orders.} 

In the remainder of this paper, we will show that the simplification (\ref{b4}) is harmless.  For KR orders, and
probably all layered causal sets, the difference between the link action and the full Benincasa-Dowker action
is negligible in the large $n$ limit.

\section{KR Counting and combinatorics \label{secKR}}

Let us start with an intuitive sketch of our argument.   Note first that for {the dominant layered causal sets,} the link term $N_0$ 
in the Benincasa-Dowker action  (\ref{b1}) goes as $n^2$ for large $n$ \cite{pst}.  In the calculations of \cite{Loomis} 
and \cite{Surya}, subleading terms of order $n$ in the action were discarded.  Such terms could perhaps be treated 
perturbatively at large $n$, but they should not be relevant for the question at hand.  In order for the $N_1$, $N_2$, 
and $N_3$ terms in the Benincasa-Dowker action to not be similarly negligible, they, too, must grow as $n^2$.  

For causal sets obtained from sprinklings in flat spacetime, it is indeed the case that
the ratio $N_J/N_0$ remains finite as  $n$ goes to infinity \cite{Glaser}, so the $N_J$ in the Benincasa-Dowker action
remain comparable.  
Layered sets, though, behave 
differently.   Figure (2a) shows a one-element interval $I(x,y)$ in a three-layer set, which contributes 
to $N_1$.  As figure (2b) shows, though, if there is a second path from $x$ to $y$,  $I(x,y)$ becomes a two-element 
interval, which contributes to $N_2$ but \emph{not} to $N_1$.  In figure (2c), the addition of a third path changes 
$I(x,y)$  to a three-element interval, contributing only to $N_3$.  For small causal sets, this is not an issue.  But 
for large sets, with many points in intermediate layers, typical pairs $x\prec y$ are connected by a great many paths, 
so $N_J$ is small for small $J$.

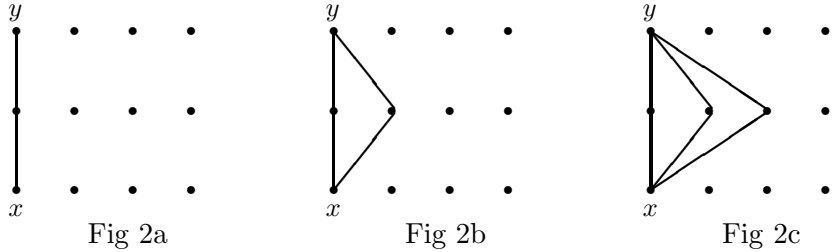
\begin{center}
\begin{picture}(320,90)(0,0)
\thicklines
\put(10,20){\circle*{3}}
\put(10,50){\circle*{3}}
\put(10,80){\circle*{3}}
\put(10,20){\line(0,1){60}}
\put(32,20){\circle*{3}}
\put(32,50){\circle*{3}}
\put(32,80){\circle*{3}}
\put(54,20){\circle*{3}}
\put(54,50){\circle*{3}}
\put(54,80){\circle*{3}}
\put(76,20){\circle*{3}}
\put(76,50){\circle*{3}}
\put(76,80){\circle*{3}}
\put(7,10){\small $x$}
\put(7,86){\small $y$}
\put(37,0){Fig 2a}
\put(130,20){\circle*{3}}
\put(130,50){\circle*{3}}
\put(130,80){\circle*{3}}
\put(130,20){\line(0,1){60}}
\put(152,20){\circle*{3}}
\put(152,50){\circle*{3}}
\put(152,80){\circle*{3}}
\put(174,20){\circle*{3}}
\put(174,50){\circle*{3}}
\put(174,80){\circle*{3}}
\put(196,20){\circle*{3}}
\put(196,50){\circle*{3}}
\put(196,80){\circle*{3}}
%\put(130,20){\line(3,2){44}}
%\put(130,80){\line(3,-2){44}}
\put(130,20){\line(4,5){23}}
\put(130,80){\line(4,-5){23}}
\put(127,10){\small $x$}
\put(127,86){\small $y$}
\put(157,0){Fig 2b}
\put(250,20){\circle*{3}}
\put(250,50){\circle*{3}}
\put(250,80){\circle*{3}}
\put(250,20){\line(0,1){60}}
\put(272,20){\circle*{3}}
\put(272,50){\circle*{3}}
\put(272,80){\circle*{3}}
\put(294,20){\circle*{3}}
\put(294,50){\circle*{3}}
\put(294,80){\circle*{3}}
\put(316,20){\circle*{3}}
\put(316,50){\circle*{3}}
\put(316,80){\circle*{3}}
\put(250,20){\line(3,2){44}}
\put(250,80){\line(3,-2){44}}
\put(250,20){\line(4,5){23}}
\put(250,80){\line(4,-5){23}}
\put(247,10){\small $x$}
\put(247,86){\small $y$}
\put(277,0){Fig 2c}
\end{picture}
\end{center}

\subsection{The link matrix}

To make this argument quantitative, we need one more ingredient.  The link matrix $\mathcal{L}$
of a causal set with elements $x_i$ is essentially a directed adjacency matrix: $\mathcal{L}_{ij}=1$ if $I(x_i,x_j)$ 
is a link and $x_i\prec x_j$, and zero otherwise.  For a three-layer set, with layers $L_1,L_2,L_3$,
the matrix $\mathcal{L}$ has the structure
\begin{align}
\mathcal{L} = \,
\begin{blockarray}{cccc}
 \scriptstyle{L_1} & \scriptstyle{L_2} & \scriptstyle{L_3}\\
 \begin{block}{(ccc)c}
 0 & P & Q & \scriptstyle{L_1}\\
 0 & 0 & R & \scriptstyle{L_2}\\
 0 & 0 & 0 & \scriptstyle{L_3}\\
\end{block}
\end{blockarray} , \qquad
\mathcal{L}^2 = \,
\begin{blockarray}{cccc}
 \scriptstyle{L_1} & \scriptstyle{L_2} & \scriptstyle{L_3}\\
 \begin{block}{(ccc)c}
 0 & 0 & PR\hspace*{2pt} & \scriptstyle{L_1}\\
 0 & 0 & 0 & \scriptstyle{L_2}\\
 0 & 0 & 0 & \scriptstyle{L_3}\\
\end{block}
\end{blockarray} 
\label{c1}
\end{align}

In general, the $(ij)$ entry in  {$\mathcal{L}^r$} is equal to the number of directed  {$r$}-step walks from 
$x_i$ to $x_j$.  For {a} three-layer causal set, such a walk is necessarily either a link or a two-step walk{, so  $r \leq 2$}.  Each two-step 
directed walk between $x_i$ in layer 1 and $x_j$ in layer 3 must include a point in layer 2, which is then a point in the
order interval $I(x_i,x_j)$.  In figure (2a), for instance, the single directed walk from $x$ to $y$ corresponds to a
one-element order interval; in figure (2b), the two directed walks correspond to a two-element interval.  In general, 
for $J>0$, the number $N_J$ of $J$-element order intervals---the number that appears in the Benincasa-Dowker
action---is equal to the number of entries in $\mathcal{L}^2$ for which $(\mathcal{L}^2)_{ij} = J$.

Our strategy for computing $N_J$ will be to find {the}  probability $Pr(N_J=K)$ for {a} random three-layer causal set,
and then to use the fact that the total number of such sets goes as $2^{n^2/4}$.  We will assume that the number
of points in each layer is proportional to $n$---that is, that layers don't disappear as $n$ becomes large---and that
there are fixed probabilities for links between layer 1 and layer 2, and  between layer 2 and layer 3.  Specifically,
denoting by $|L_i|$ the number of points in layer $i$, {let}  
\begin{align}
&|L_1| = \gamma_1n,\ |L_2| = \gamma_2n,\ |L_3| = \gamma_3n\quad\hbox{(with $\gamma_1+\gamma_2+\gamma_3=1$)}\\
&\hbox{Probability $p$ of a link from a point in $L_1$ to a point in $L_2$}\\
&\hbox{Probability $q$ of a link from a point in $L_2$ to a point in $L_3$} 
\label{c2}
\end{align}

Next look more carefully at the link matrix (\ref{c1}).  In $\mathcal{L}^2$, $PR$ is an $|L_1|\times|L_3|$ matrix, 
whose $(ij)$ entry is a dot product of two vectors of length $|L_2| =\gamma_2n$, namely ${\bf P}_{(i)} =  P_{(i)k}$ and 
${\bf R}_{(j)} =R_{k(j)}$, where $k$ designates the component while $i$ and $j$ are treated as fixed labels.  So the 
problem is to 
\begin{itemize}
\item[(i)] compute the probability $P_J$ that this dot product is $J$ for a random causal set;
\item[(ii)] compute the probability $Pr(N_J=K)$ that exactly $K$ such products appear in $PR$;
\item[(iii)] find the number of causal sets with $N_J=K$, and see how it behaves in the large $n$ limit.
\end{itemize}

\subsection{Computing $P_J$}

We are looking at the product of the vectors ${\bf P}_{(i)} =  P_{(i)k}$ and ${\bf R}_{(j)} =R_{k(j)}$.  Now,
$P_{ik}$ is $1$ if {$e_i$} in layer 1 links to {$e_k$} in layer 2, which occurs with a probability $p$, while
$R_{kj}$ is $1$ if {$e_k$} in layer 2 links to {$e_j$} in layer 3, which occurs with a probability $q$.
Thus the probability that a \emph{fixed} set of $J$ {components in ${\bf P_{(i)}}\cdot {\bf R_{(j)}}$ }
 have value $1$ and 
the remainder have value $0$ is
$$(pq)^J(1-pq)^{\gamma_2n-J}.$$  But there are $\displaystyle \gamma_2n\choose J$ ways of choosing these $J$ components,
so 
\begin{align}
P_J = {\gamma_2n\choose J} (pq)^J(1-pq)^{\gamma_2n-J}
\label{ca1}
\end{align}
Note that the leading behavior of $P_J$ for fixed $J$ and large $n$ is
\begin{align}
P_J \sim A_J\,n^{J}{\alpha}^{\gamma_2n}
\label{ca2}
\end{align}
with {$\alpha=1-pq<1$}, so for fixed $J$,  $P_J\rightarrow0$ as $n\rightarrow\infty$.

\subsection{Computing $Pr(N_J=K)$}

The matrix $PR$ has $|L_1|\times|L_3| = \gamma_1\gamma_3n^2$ entries, each having a probability 
$P_J$ of equaling $J$.  Thus the probability that a  \emph{fixed} set of $K$ entries have value $J$ and the 
remainder do not is
$$P_J{}^K(1-P_J)^{\gamma_1\gamma_3n^2-K}$$
But there are $\displaystyle \gamma_1\gamma_3n^2\choose K$ ways of choosing such entries.  Hence
\begin{align}
Pr(N_J=K) = {\gamma_1\gamma_3n^2\choose K}P_J{}^K(1-P_J)^{\gamma_1\gamma_3n^2-K}
\label{cb1}
\end{align}

\subsection{Large $n$ asymptotics}

For a three-layer set, the maximum value of $N_1$ is achieved when there is exactly one
path between each point in layer 1 and each point in layer 3.  Since there are $\gamma_1n$ points
in layer 1 and $\gamma_3n$ points in layer 3, $N_1^{\rm max} = \gamma_1\gamma_3n^2$.  
$N_2^{\rm max}$ and $N_3^{\rm max}$ take the same value, although
the maxima can't be reached simultaneously.  Since we are interested in the question of whether
the $N_J$ grow as $n^2$, let us take
\begin{align}
K = {s} N_J^{\rm max} ={ s}  \gamma_1\gamma_3n^2
\label{cc0}
\end{align}
and look at the large $n$ behavior of $Pr(N_J=K)$.

Begin with the last term in (\ref{cb1}).  For large $n$, $P_J$ is exponentially small, so by 
(\ref{ca2}),
\begin{align}
\ln (1-P_J)^{\gamma_1\gamma_3n^2-K} \sim -(1-s)\gamma_1\gamma_3n^2 P_J
   \sim -(1-s)\gamma_1\gamma_3A_J\,n^{J+2}\alpha^{\gamma_2n}
\label{cc1}
\end{align}
which goes to zero exponentially for large $n$.  Hence
\begin{align}
(1-P_J)^{\gamma_1\gamma_3n^2-K} \sim 1 - \hbox{\it exponentially small corrections}
\label{cc2}
\end{align}
Inserting back into (\ref{cb1}),
\begin{align}
Pr(N_J=K) \sim {\gamma_1\gamma_3n^2\choose s\gamma_1\gamma_3n^2}
    \left(A_J\,n^{J}\alpha^{\gamma_2n}\right)^{s\gamma_1\gamma_3n^2}
\label{cc6}
\end{align}
The binomial coefficient in (\ref{cc6}) is maximum when $s=\frac{1}{2}$, where
$${\gamma_1\gamma_3n^2\choose s\gamma_1\gamma_3n^2} \sim 2^{\gamma_1\gamma_3n^2}$$
But the term that follows has a leading behavior
$$ \alpha^{s\gamma_1\gamma_2\gamma_3n^3} = 2^{-s\gamma_1\gamma_2\gamma_3|\log_2 \alpha|n^3}$$
which clearly dominates for large $n$.  Thus the asymptotic behavior is
\begin{align}
Pr(N_J= {sN_{J}^{\mathrm{max}}}) \sim 2^{-s\gamma_1\gamma_2\gamma_3|\log_2\alpha|n^3}
\label{cc7}
\end{align}
 
\subsection{Implications for the action}

The Benincasa-Dowker action (\ref{b1}) and the link action (\ref{b4}) differ by terms
proportional to $N_1$, $N_2$, and $N_3$.  For KR orders, or more generally three-layer causal sets, 
we can now answer the question of whether this difference matters.
As noted earlier,  the leading term in both actions is proportional to $N_0$, 
which for layered causal sets goes as $n^2$ for large $n$.  For the $N_1$, $N_2$, and $N_3$ terms
in the Benincasa-Dowker action to give more than subleading corrections, they, too, must grow as $n^2$.   
We can now see that they do not.

Indeed, pick arbitrarily small constants $\varepsilon_J$, and ask how many three-layer
causal sets have $N_J>\varepsilon_Jn^2$.  From (\ref{cc0}) and (\ref{cc6}), the probability that
any particular causal set satisfies this condition is
\begin{align}
Pr(N_J>\varepsilon_Jn^2) 
   \sim \int\limits_{{\varepsilon_J}/{\gamma_1\gamma_3}}^1\!\!\!ds\,\,2^{-s\gamma_1\gamma_2\gamma_3|\log_2\alpha|n^3}
   \sim 2^{-\epsilon_J\gamma_2|\log_2\alpha|n^3} 
\label{cd1}
\end{align}where the integral is dominated by its lower limit.
But the total number of three-layer set with $n$  elements increases only as $2^{n^2/4}$.   Thus for large 
enough $n$, the expected number of three-layer sets that contribute more than subleading corrections to the 
Benincasa-Dowker action is zero.

We can further strengthen this result.  Let us ask a weaker question: how many three-layer sets have $N_J>\beta_Jn$
for some fixed constants $\beta_J$?  From (\ref{cb1}), 
\begin{align}
Pr(N_J=\beta_Jn) 
   = {\gamma_1\gamma_3n^2\choose \beta_Jn}P_J{}^{\beta_Jn}(1-P_J)^{\gamma_1\gamma_3n^2-\beta_Jn}
\label{cd2}
\end{align}
As before, the last term is very nearly one, while the binomial coefficient is now
$${\gamma_1\gamma_3n^2\choose \beta_Jn} \sim \left(\frac{\gamma_1\gamma_3e}{\beta_J}n\right)^{\beta_Jn}$$
For large enough $n$, the leading behavior of $Pr(N_J=\beta_Jn)$ is again controlled by the term
$$P_J{}^{\beta_Jn} \sim 2^{-\beta_J\gamma_2|\log_2\alpha|n^2}$$
Thus
\begin{align}
Pr(N_J>\beta_Jn) 
   \sim \int\limits_{\beta_J}^{\gamma_1\gamma_3n}\!\!\!d\beta'\,\,2^{-\beta'\gamma_2|\log_2\alpha|{n^2}}
   \sim 2^{-\beta_J\gamma_2|\log_2\alpha|n^2}
\label{cd3}
\end{align}
Again, the number of three-layer causal sets goes as $2^{n^2/4}$, so for large $n$, the expected number
of such sets is
\begin{align}
\#(N_J>\beta_Jn) \sim 2^{\left(\frac{1}{4} - \beta_J\gamma_2|\log_2\alpha|\right)n^2}
\label{cd4}
\end{align}
For
\begin{align}
\beta_J>\frac{1}{4\gamma_2|\log_2\alpha|}
\label{cd5}
\end{align}
this number falls to zero very quickly as $n$ becomes large.   

We conclude that for three-layer sets, including KR orders, the terms $N_{J>0}$ in the Benincasa-Dowker 
action (\ref{b1}) increase at large $n$ at most as
\begin{align}
N_J\sim\frac{n}{4\gamma_2|\log_2\alpha|}
\label{cd6}
\end{align}
This makes them subleading corrections to the link action.  The superexponential suppression of KR orders 
found in \cite{Surya} is therefore not just a feature of the link action, but holds for the full Benincasa-Dowker
action, the causal set analog of the Einstein-Hilbert action.  The gravitational path integral suppresses KR
orders.

\section{Conclusion}

We have seen that for a large range of couplings, the gravitational path sum with the discretized Einstein-Hilbert
action strongly suppresses the two largest classes of ``bad'' causal sets, the KR orders and the two-layer sets. 
An obvious question is whether these results can be generalized to the whole hierarchy of layered causal sets
described in \cite{dhar,pst}.  To show this would require two steps: a generalization of this work to justify using
the link action, {and an extension of the results of \cite{Surya} beyond the KR case.}

The first step is almost certainly possible.  While the combinatorics becomes more complicated, the qualitative
argument of section \ref{secKR} continues to hold for sets with more than three layers.  Work on the details is
in progress.  {As for the second step, 
the results of \cite{Surya} are in fact stronger than stated. For a fixed number of links the leading order
contribution to all $k$-layer orders comes from the link maximizing configurations, which are precisely the KR and
symmetric two-layer orders. Thus, even when the $k$-layers are included in the sum, they do not contribute to leading order.}

Of course, even if one can show that all layered causal sets are suppressed, this is not quite enough to show
the emergence of a continuum from the gravitational path integral.  There are certainly other non-manifoldlike
causal sets that do not have a simple layered structure.  More generally, while we have an explicit construction of
manifoldlike causal sets, we do not yet have a good intrinsic characterization of such sets.  Still, though, the
layered sets make up such an enormous share of the ``bad'' causal sets that their elimination is a major
step forward.

Finally, if causal sets are primary and the continuum is emergent, a deeper question remains.  The 
Benincasa-Dowker action (\ref{b1}) came from the continuum, starting with causal sets that were constructed 
by hand to be manifoldlike and looking for an approximation of the continuum Einstein-Hilbert action.
 {The most obvious tie to the continuum and classicality is the dimension, which we have assumed to be $4$ in the
  Benincasa-Dowker action (\ref{b1}).  In arbitrary dimensions as well,  the Benincasa-Dowker-Glaser action \cite{bdg,glaser} is
  an alternating sum over the $N_J$, where $J$
  ranges from $0$ to a dimension dependent $J_{\mathrm{max}}$.  Thus our results  show more generally that the link action suffices to
  suppress KR orders  in any dimension.} 
{ At a more fundamental level, however, what}  is really needed is a derivation of the action from first principles, entirely within the context of causal
sets.

\vspace{1.5ex}
\begin{flushleft}
\large\bf Acknowledgments
\end{flushleft}

S.~C.\ was supported in part by Department of Energy grant
DE-FG02-91ER40674.

%\bibliographystyle{utphys}
%\bibliography{Jcausets}

% \begin{thebibliography}{99}
% \bibitem{causets}
 
% \end{thebibliography}
\end{document}